\documentclass[dvips,noinfoline]{imsart}
\usepackage[french,english]{babel}
\usepackage{amsmath}
\usepackage{amssymb}
\usepackage{amsthm,graphicx}
\usepackage{natbib,epsfig}

\usepackage[T1]{fontenc}
\def\R{\mathbb{R}}
\def\P{\mathbb{P}}
\def\E{\mathbb{E}}
\def\R{\mathbb{R}}
\def\L{\mathbb{L}}
\def\1{\ \mbox{\large I}}
\def\var{\mbox{var}\,}
\def\cov{\mbox{cov}\,}
\newtheorem{prop}{Proposition}

\newtheorem{theo}{Theorem}
\newtheorem{lem}{Lemma}

\begin{document}

\begin{frontmatter}
\begin{aug}

\title{Testing randomness \\of spatial point
patterns\\ with the Ripley statistic}
\runtitle{Testing randomness of point patterns}
\author{\fnms{Gabriel} \snm{Lang*}\ead[label=e2]{gabriel.lang@agroparistech.fr}}
\address{UMR 518 Math\'{e}matique et Informatique appliqu\'{e}es, \\AgroParisTech,
\\19 avenue du Maine,\\75732 PARIS CEDEX 15, France.\\\printead{e2}}
\author{\fnms{Eric} \snm{Marcon}\ead[label=e1]{eric.marcon@ecofog.gf}}
\address{UMR 745 Ecologie des For\^{e}ts de Guyane,\\
AgroParisTech,\\ Campus agronomique BP 316,\\ 97379 KOUROU CEDEX,
France.\\\printead{e1}} \affiliation{ AgroParisTech}
\runauthor{G. Lang and E. Marcon}
\end{aug}

\begin{abstract}
Aggregation  patterns are often visually detected in sets of
location data. These clusters may be the result of interesting
dynamics  or the effect of pure randomness. We build an
asymptotically Gaussian test for the hypothesis of randomness
corresponding to a  Poisson point process. We first compute the
exact first and second moment of the Ripley K-statistic under the
homogeneous Poisson point process model. Then we prove the
asymptotic normality of a vector of such statistics for different
scales and compute its covariance matrix. From these results, we
derive a test statistic that is chi-square distributed. By a
Monte-Carlo study, we check that the test is numerically tractable
even for large data sets and also correct when only a hundred of
points are observed.
\end{abstract}



\end{frontmatter}

\section{Introduction}Analysis of point patterns is relevant in many sciences: cell biology, ecology or spatial economics.
 The observation of clusters in point locations  is considered as a hint for non observable dynamics.
 For example the clustering of tree locations in a forest may come from better soil conditions or
 from spreading of seeds of a same mature individual; but clusters are also observed in random
 distribution as a Poisson point process sample. It is therefore essential to
 distinguish between clusters resulting from relevant interactions
 or from complete randomness.
 \cite{RI76,RI77}--- is a widely used tool to quantify the structure of
point patterns, especially  in ecology, and is well referenced in
handbooks \citep{RI81,DI,STKEME,CR,MOWA,IL}. Up to a renormalization by the intensity of the process, this
statistic denoted here $\hat K(r)$ estimates the expectation $K(r)$
of the number of neighbors at distance less than $r$ of a point in
the sample. The observed $\hat K(r)$ is compared to the value of $K(r)$ for a homogeneous Poisson point
process with the same intensity as the data, chosen as a null hypothesis:
 the  Poisson point process is characterized by an independence
of  point locations, modelling  an
 absence of interactions between individuals in ecosystems. In this case
 $K(r)$ is simply the mean number of
points in a ball of radius $r$ divided by the intensity, that is $\pi  r^2$.
If $\hat K(r)$ is significantly larger than $\pi  r^2$ (respectively smaller),
the process is considered
as aggregated (respectively over-dispersed) at distance $r$. \\
To decide if the difference is statistically significant, we build
a test of the Poisson process hypothesis; we need to know the
distribution of $\hat K(r)$ for this process. But even the
variance is not known and statistical methods generally rely on
Monte-Carlo simulations. \cite{RI79} used them to get confidence
intervals. Starting from previous results \citep{SAFU}, he also gave critical values for the $L$
function, a normalized version of $K$ introduced by
\cite{BES}. These critical values are valid
asymptotically, for a large number of points but low intensity, so
that both edge effects and point-pair dependence can be neglected.
Further computations of confidence interval bands based on
simulation have been proposed in \cite{KO} and corrected in
\cite{CH}. But the simulation is a practical issue for large point
patterns, because computation time is roughly proportional to the
square of the number of points (one has to calculate the distances
between all pairs of points) multiplied by the number of
simulations.
\\
We propose here to compute the exact variance of the Ripley
statistic. \cite{WAFE} studied this variance. But they ignored
that point pairs are not independent even though points are (eq.
A8, p. 235), thus their derivation of the variance of $\hat K(r)$ was
erroneous. The right way to compute the covariance is to consider
that it is a $U$-statistic as remarked in \cite{RI79}, then to use
the H\"{o}ffding decomposition. As the variance is not enough to
build a test, we study the distribution of the statistic. We prove
its asymptotic normality as the size of the observation window
grows. It is then easy to build an asymptotically Gaussian test.
\\
Another concern is to test simultaneously the aggregation/dispersion at different scales.
This is rarely correctly achieved in  practical
computations with Monte-Carlo simulations. The confidence bands or
test rejection zone are often determined without taking the dependence
between the numbers of neighbors at different scales into account.
As an exception \cite{DUOV} provide a heuristic multiscale test. In our main theorem, we
consider a set of scales $(r_1,\ldots,r_d)$, compute the covariance
matrix of the $K(r_i)$ and prove the asymptotic normality for the
vector $(K(r_1),\ldots,K(r_d))$. From this we propose the first rigorous  multiscale
test of randomness for point patterns.
\\
The paper is built as follows: Section 2 introduces the precise definition of $K(r)$ and the
current definition of $\hat K(r)$. In Section 3, after the
definition  of our statistics (no edge-effects correction, known or
unknown intensity), we list the main results of the paper: exact
bias due to the edge effects and exact variance of $\hat K(r)$ for a
homogeneous Poisson process with known or unknown intensity;
covariance between $\hat K(r)$ and $\hat K(r')$ for two different
distances $r$ and $r'$. The main theorem contains the convergence of
the vector $(K(r_1),\ldots,K(r_d))$ to a Gaussian distribution with
explicit covariance in the following asymptotic framework: data from
the same process are collected on growing squares of observation.
These results allow a simple, multiscale and efficient test
procedure of the Poisson process hypothesis. Section 4 provides a
Monte-Carlo study of the test and Section 5 gives our conclusions. The last section contains the proofs.
Technical integration lemmas are postponed in the appendix.

\section{Definition of the Ripley $K$-function}
We recall the  characterizations of the dependence of the locations
for a general point process $X$ over $\R^2$. We refer to the presentation of \cite{MOWA}.
\subsection{Definitions} For a point process $X$, define the point process
$X^{(2)}$ on $\R^2\times\R^2$ of all the couples of two different
points of the original process. The intensity of this new process
gives information on the simultaneous  presence of points in the
original process. Denote $\rho^{(2)}(x,y)$ its density (called the
second-order product density). The  Poisson process of density
$\rho(x)$ is such that
$\rho^{(2)}(x,y)=\rho(x)\rho(y)$.\\
 The Ripley statistic is a way
to estimate the density $\rho^{(2)}(x,y)$. Precisely it is an
estimate of the integral on test sets of the ratio
$g(x,y)=\rho^{(2)}(x,y)/\rho(x)\rho(y)$. The function $g(x,y)$
characterizes the fact that the points $x$ and $y$ appear
simultaneously in the samples of $X$. If $g(x,y)=1$, the points
appear independently. If $g(x,y)<1$, they tend to exclude each
other; if $g(x,y)>1$, they appear more frequently together. \\
We assume the translation invariance of the point process:
$g(x,y)=g(x-y)$. In order to estimate the function $g$, we define
its integral as the set function ${\cal K}$. Let $A$ be a Borel
set: $${\cal K}(A)=\int_Ag(x)dx.$$ If we also assume that the
point process is isotropic, we define the Ripley $K$-function as
$$
K(r)={\cal K}(B(x,r)),
$$
where $B(x,r)$ is the closed ball with center $x$ and radius $r$.
The translation invariance implies that ${\cal K}(B(x,r))$ does not depend on $x$.
For example, if the process is a Poisson process then $g(x)=1$ and
$K(r)=\pi r^2$. We define the Ripley statistic that estimates the
$K$-function. Let $A$ be a bounded Borel set of the plane $\R^2$, $m$ the
Lebesgue measure and $\widehat\rho$  an estimator of the local intensity of the process; for a realization $ S$ of the point process $X$,
$S=\{X_1,\ldots,X_N\}$, the Ripley statistic is defined by
$$
\widehat K_A(r)=\frac 1 {m(A) }\sum_{X_i\neq X_j \in S}\frac{
\1\{d(X_i,X_j)\leq r\}}{ \widehat\rho~\!(X_i)~\widehat\rho~\!(X_j)}.
$$

\section{Main results}
This section presents the theoretical results on the Ripley statistic and the resulting test.
\subsection{Definitions}Throughout the paper, we refer to the indicator function $\1$, the expectation $e_{r,n}$, the centred indicator function $h$ and its conditional expectation $h_1$. We gather here these definitions.\\
Let $n$ be an integer; $A_n$ denotes the square $[0,n]^2$; $U$ is
a random location in $A_n$ with an uniform random distribution;
its density is $1/n^2$ with respect to the Lebesgue measure
$d\xi_1d\xi_2$ over $A_n$. $V$ is a random location with the same
distribution as $U$ and independent of $U$. We denote $d(x,y)$ the
Euclidean distance between $x$ and $y$ in the plane, and $\1
\{A\}$ the indicator function of set $A$. We define
$e_{r,n}=\E(\1\{d(U,V)\leq r\})$,  $ h(x,y,r)=\1\, \{d(x,y)\leq
r\}-e_{r,n}$ and $h_1(x,r)=\E (h(U,V,r)|\ V=x)$.
\subsection{Assumptions} We assume that  $X$ is a
homogeneous Poisson process on $\R^2$ with intensity $\rho$.
 We consider that the
data are available on the square $A_n$. $S=\{X_1, \ldots,X_N\}$ is
the  sample of observed points. We consider two cases:
\begin{enumerate}
    \item If the intensity $\rho$ is known, the Ripley statistic is expressed as
$$
\widehat K_{1,n}(r)=\frac 1 {n^2\rho^2 }\sum_{X_i\neq X_j \in S}
\1\{d(X_i,X_j)\leq r\}.
$$
    \item If the intensity $\rho$ is unknown, we choose
to estimate $\rho^2$ by the unbiased estimator $\displaystyle\widehat
{\rho^2}=N(N-1)/n^4$ \citep{STST} and define
$$
\widehat K_{2,n}(r)=\frac {n^2} {N(N-1) }\sum_{X_i\neq X_j \in S}
\1\{d(X_i,X_j)\leq r\}.
$$
\end{enumerate}
\subsection{Bias}
It is known that a large number of neighbors of the points located
near the edges of $A_n$ may lie outside $A_n$ causing a bias in
the estimation. We compute the bias due to this edge effect.
\begin{prop}\label{prbi}
Assume that $r/n<1/2$.
\begin{eqnarray*}
 \E\widehat
K_{1,n}(r)-K(r)&=&r^2\left( -\frac {8r}{3n}+\frac
{r^2}{2n^2}\right).\\
 \E\widehat K_{2,n}(r)-K(r)&=&r^2\left(
-\frac {8r}{3n}+\frac {r^2}{2n^2}\right)\\&&-r^2e^{-\rho
n^2}\left(\pi -\frac {8r}{3n}+\frac {r^2}{2n^2}\right)\left(1+\rho
n^2e^{-\rho n^2}\right).
\end{eqnarray*}
\end{prop}
{\it Notes:}
\begin{itemize}
    \item   The assumption that $r/n$ is less than $1/2$
means that at least some balls of radius $r$  are included in the
square $A_n$.
    \item The additional term for $K_{2,n}$ corresponds to the
probability to draw a sample with zero or one point in the square.
This
probability is so low that the term gives a zero contribution as soon as the mean number of points $\rho n^2$ is larger than $20$.
    \item The proof may be adapted for a convex polygon of perimeter $Ln$ to
compute the first order term of the bias; for $u=1$ or $2$:
$$\E\widehat K_{u,n}(r)-K(r)=-\frac {2Lr^2}3\frac {r}{n}+O\left(\frac {r^2}{n^2}\right).$$
\end{itemize}

\subsection{Variance}
We compute the  covariance matrix  of $\widehat K_{u,n}(r)$ for $u=1$ or 2. We
get an exact computation for the variance, that can be used for
any value of $n$.
 \begin{prop}\label{pvar}For $0<r<r'$,
\begin{eqnarray*}
\var(\widehat K_{1,n}(r))
  &=&\frac {2e_{r,n}}{\rho^2}+\frac {4n^2e_{r,n}^2}{\rho}+\frac
  {4n^2}{\rho}\E h_1^2(U,r),\\
  \cov(\widehat K_{1,n}(r),\widehat
K_{1,n}(r'))&=&\frac {2e_{r,n}}{\rho^2}+\frac
{4n^2e_{r',n}e_{r,n}}{\rho}+\frac
  {4n^2}{\rho}\cov(h_1(U,r'),h_1(U,r)),\end{eqnarray*}
\begin{eqnarray*}\var(\widehat K_{2,n}(r))
&=&2n^4\E\left(\frac{\1\{N>1\}}{N(N-1)}
\right)\left(e_{r,n}-e_{r,n}^2\right)
\\&+&4n^4\E\left(\frac{\1\{N>1\}(N-2)}{N(N-1)}\right)\E h_1^2(U,r)\\&+&n^4e^{-\rho n^2}\left(1+\rho
n^2\right)\left(1-e^{-\rho n^2}-\rho n^2e^{-\rho
n^2}\right)e^2_{r,n},\\
\cov(\widehat K_{2,n}(r),\widehat
K_{2,n}(r'))&=&2n^4\E\left(\frac{\1\{N>1\}}{N(N-1)}
\right)\left(e_{r,n}-e_{r',n}e_{r,n}\right)\\ &+&
4n^4\E\left(\frac{\1\{N>1\}(N-2)}{N(N-1)}\right)\cov(h_1(U,r'),h_1(U,r))\\&+&n^4e^{-\rho
n^2}\left(1+\rho n^2\right)\left(1\!\!-e^{-\rho n^2}\!\!-\rho
n^2e^{-\rho n^2}\right)e_{r',n}e_{r,n},\end{eqnarray*} where
\begin{eqnarray*}
  e_{r,n} &=& \frac{\pi r^2}{n^2} -\frac {8r^3}{3n^3}+\frac {r^4}{2n^4}, \\
\E h_1^2(U,r)
        &=&\frac{r^5}{n^5}\left(\frac 8 3\,\pi -{\frac {256}{45}}\right)
        +\frac{r^6}{n^6}\left(\frac {11}{48}\,\pi -\frac {56}
{9}\right)+\frac 8 3\frac{r^7}{n^7}-\frac 1 4\frac{r^8}{n^8}.
\end{eqnarray*}
\end{prop}
{\it Notes}:
\begin{itemize}
    \item The variances of both estimators are exact and can be
computed at any precision, as inverse moments of the Poisson
variable correspond to fast converging series. But  these series
may be difficult to evaluate with mathematical softwares, because of the large
value of the Poisson parameter.
 \item The covariances are not
explicit because  the terms  $\cov(h_1^2(U,r'),h_1^2(U,r))$
involve terms that have to be numerically integrated. \item The
leading terms of the variances of $K_{1,n}(r)$ and $K_{2,n}(r)$ as
$n$ tends to infinity  are $2\pi r^2/n^2\rho^2+4\pi r^4/n^2\rho$
and $2\pi r^2/n^2\rho^2$.
\end{itemize}

\subsection{Central Limit Theorem}
We show that a normalized vector of Ripley statistics for
different $r$ converges in distribution to a normal vector. Let ${\cal N}(0,\Sigma)$ denote the Gaussian multivariate centred distribution with covariance matrix   $\Sigma$.
\begin{theo}\label{tcl} Let $d$ be an integer, $0<r_1< \ldots< r_d$ a set of
 reals and define ${\cal K}_{u,n}=(\widehat K_{u,n}(r_1),
\ldots ,\widehat K_{u,n}(r_d))$. Then $ n\sqrt{\rho}({\cal
K}_{u,n}-\pi (r_1^2, \ldots r_d^2))$ converges in distribution to
${\cal N}(0,\Sigma)$ as $n$ tends to infinity, where for $s$ and
$t$ in $\{1,\ldots,d\}$\begin{itemize}
    \item if $u=1$, $\displaystyle
      \Sigma_{s,t}=\frac{2\pi (r_s^2\wedge
      r_t^2)}{\rho}+{4\pi^2r_s^2r_t^2}$.
    \item if $u=2$, $\displaystyle
   \Sigma_{s,t}=\frac{2\pi (r_s^2\wedge r_t^2)}{\rho}.$
 \end{itemize}

 \end{theo}
{\it  Note}: The first term of the variance corresponds to a
situation where the couples of points are independent from each
others; this was used as an approximation without proof in
\cite{WAFE}; our work proves that the actual variance and limit
process are different in the first case and that the approximation
holds only in the second case.
\subsection{Applications to test statistics}
From Theorem \ref{tcl}, we deduce that $T_u=\Sigma^{-1/2}{\cal
K}_{u,n}$ is asymptotically ${\cal N}(0,I_d)$ distributed. For the
hypothesis
\begin{center}
$H_0$: $X$ is a homogeneous Poisson process of intensity $\rho$
\end{center} we use $T^2=\|T_u\|_2^2$ as a test statistic with rejection
zone for the level $\alpha$:  $$T^2>\chi_\alpha^2(d).$$ where
$\chi_\alpha^2(d)$ is the $(1-\alpha)$-quantile of the $\chi^2(d)$
distribution.
 \\{\it Note}: the covariance matrix $\Sigma$ depends on the
intensity parameter $\rho$, so that in the case of the unknown
parameter we have to use an estimate of $\rho$ in the formula
defining $\Sigma$.
\section{Simulations}
We study the empirical variance of the proposed statistics by a
Monte-Carlo simulation. Then we apply
the test procedure to simulated data sets, observe the number of
rejections and compare it to the level of the test.

\subsection{Variance}
\begin{figure}
\centering
\makebox{\includegraphics{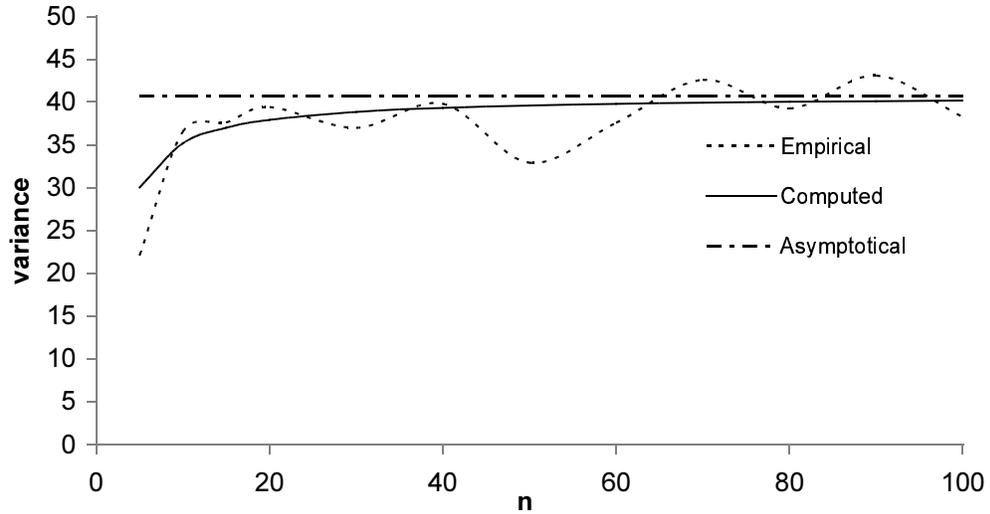}}\caption{\label{f1}Comparison
of normalized variances for $K_1(1)$, $\rho=5$}
\end{figure}

We simulate a sample of 1000 repetitions with $\rho=5$ and compare
(after renormalization by $n\sqrt \rho$) the empirical variance and
the exact computed variance with the limit variance for different
value of $n$ (figure \ref{f1}).
With 1000 repetitions, the oscillations of the empirical variance
are still large; we will use a larger number of
repetitions in the following study of the test.\\
The convergence of the computed variance to the limit value is not
so fast and for applications with hundreds of points (corresponding
in figure \ref{f1} to $n<15$) the distance between the variances is
still large. A preliminary study, not presented here, showed that
the test procedure is perturbed by an small error in the covariance
matrix, as  we tried simplified versions of the covariance by
bounding or ignoring the corner contribution $C(A_n^{3,3})$ (see in
the proof section). It is crucial to use an accurate computation of
the covariance matrix to have a correct approximation of the square
root inverse matrix $\Sigma^{-1/2}$. Therefore we will use the exact
formula instead of the asymptotic formula in the test procedure.\\

\subsection{Test}
In the known parameter case, the computation of the test statistic
$T_1$ is straightforward; we also build a statistic $T^*_1$ using
the empirical covariance matrix of the sample. The advantage of
$T^*_1$ is that it is orthogonal by construction and should lead
to better results. But the covariance matrix is not observable
when we dispose of one sample, so that the test procedure based on
$T^*_1$ is unfeasible. It is an idealized
version, used to compare the corresponding number of
rejections. To avoid the statistical dependence between the sample
and the estimator of the covariance matrix, we also build a
statistic $T'_1$ where we generate a additional independent sample
of the Poisson process with intensity $\rho$
to compute the  empirical covariance matrix.\\
In the unknown parameter case, the computation of the test
statistic $T_2$ is similar. In the variance formula the unknown
parameter  $\rho$ is replaced by the estimator $N/n^2$. We also
choose to replace the expectation
$\E\left(\1\{N>1\}/(N(N-1))\right)$ by the observed value
$1/(N(N-1))$ and $\E\left(\1\{N>1\}(N-2)/(N(N-1))\right)$ by
$(N-2)/(N(N-1))$, because the dispersion of a Poisson variable is
low with respect to the expectation when its intensity is large.
The construction of  $T^*_2 $ is the same as for $T^*_1$. The case
of $T'_2$ is not studied because, as $\rho$ is unknown, one would
have to generate
an additional sample for each estimated value of $\rho$.\\
The test output is a Bernoulli random variable with parameter
$\alpha$. With a sufficient index of repetition $m$, the mean
number of rejection is close  to a normal variable with
expectation $\alpha$ and variance $\alpha(1-\alpha)/m$. We
consider that the test works when the observed frequency of
rejection is in the 95\% Gaussian  confidence  interval
$[\alpha-1.96\sqrt{\alpha(1-\alpha)/m},\alpha+1.96\sqrt{\alpha(1-\alpha)/m}]$.
With $m=10000$ and $\alpha=0.05$, the interval is
$[0.0457;0.0543]$ so that the percentile of rejection in table
\ref{tab01} should lie in $[4.57;5.43]$. Stars indicate the values
outside the confidence interval.

\begin{table}
\caption{\label{tab01}Percentile of rejection over $10000$
repetitions of the test with level $\alpha=0.05$.}  \centering
  \fbox{
 \centering
\begin{tabular}{l l l|l|l|l|l|l}
\multicolumn {3}{c|}{Poisson} &$T^*_1$&$T'_1$&$T_1$&  $T^*_2$& $T_2$\\
  \hline
$n=30$&$\rho=1$&$r=(1,2,5)$&5.40&5.04&5.20&5.01&5.10\\
$n=10$&$\rho=5$&$r=(1,2,5)$&$5.61^*$&5.40&5.19&5.38&5.37\\
$n=10$&$\rho=5$&$r=(1,2,\ldots,10)$&5.13&5.32&$5.76^*$&$6.67^*$&$6.01^*$\\
$n=10$&$\rho=1$&$r=(1,2,5)$&$5.67^*$&$5.86^*$&$5.81^*$&5.30&5.25\\
$n=10$&$\rho=.5$&$r=(1,2,5)$&$5.52^*$&$5.73^*$&$5.52^*$&$5.60^*$&4.91\\
$n=10$&$\rho=.2$&$r=(1,2,5)$&$6.40^*$&$6.84^*$&$6.59^*$&$6.59^*$&5.22\\
 \end{tabular} }
\end{table}
\noindent The performances in the case of a known parameter ($T_1$, $T^*_1$
and $T'_1$) are good except when the number of points is small. The
unfeasible tests $T^*_1$ and $T'_1$ based on the empirical
covariance have no better performance than the test $T_1$. The error
of the empirical covariance is probably still to large.  The only
exception is the third line where a large number
of values of $r$ are considered simultaneously. \\
The test $T_2$ performs better than $T_1$  for small data sets.
The only exception is the case of a large number of scales. The
poor performance of $T_1$ and $T_2$ in this case may result from
numerical instabilities in the covariance matrix inversion as its
dimension is larger. The departure from normality may also be
larger in this case (some classes of inter-point distances being
weakly represented in the sample).
 With this exception, the test based on $T_2$ works perfectly.\\In table 2, we
investigate the power of the test $T_2$ by simulating two
Thomas cluster processes \citep{TH}. A Thomas
process is a Neyman-Scott process; the germs of the clusters are
drawn as a sample of a homogeneous Poisson process of intensity
$\kappa$ . For each germ, an inhomogeneous Poisson process is
drawn with intensity measure $\mu f$, where $f$ is the density of
the Gaussian two-dimensional vector centered on the germ and with
independent coordinates of variance $\sigma$. The Thomas process
results from the superposition of these Poisson processes. The germs
are not conserved. The parameters of the two processes are such
that clusters are not visually detectable in the first process and
evident in the second one.
\begin{table}
\caption{\label{tab02}Percentile of rejection over $10000$
repetitions of the test with level $\alpha=0.05$.}  \centering
  \fbox{
 \centering
\begin{tabular}{l l l|l}
\multicolumn {3}{c|}{Thomas}& $T_2$\\
 \hline
 $n=10$&$(\kappa,\mu,\sigma)=(1,5,3)$&$r=(1,2,5)$&$71.6$\\
$n=10$&$(\kappa,\mu,\sigma)=(0.5,10,0.5)$&$r=(1,2,5)$&$100$\\
 \end{tabular} }
\end{table}
The
test rejects 71\% of the first sample and systematically the
second one. The  test is more powerful than a visual observation
of the data, detecting invisible clusters. A rigorous analysis of
the distribution of the statistic for dependent point process
models should allow to conclude on the power of our test but such
a study is beyond the scope of this paper.
\section{Conclusion}
We provide an efficient test of the null hypothesis of a
homogeneous Poisson process for  point patterns in a square
domain. This is a theoretical and practical improvement on
preexisting methods: Monte-Carlo simulations are untractable when
the number of points increases. With a personal computer,
calculating $K$ for 10,000 simulations of a 10,000-point set is
not feasible (or it will take months). \cite{MAPU} applied $K$ to
a 36,000-point data set (the largest ever published as far as we
know), but had to
limit the number of simulations to 20. \\
We suggest to change the treatment of edge effects. Instead of correcting edge effect on each sample to
reduce the bias, we compute the exact bias. The use of sample
correction (for each point of the data) has not been questioned
since  Ripley's original paper, except by \cite{WAFE}.\\
We also point out that the test can be used on samples with a
few dozens of points as encountered in actual data sets. It works
correctly with such small data sets, even if it is based on
asymptotic normality. This is due to the fact that the bias and
variance are known exactly and not asymptotically; the non-normality of the statistics for small data sets seems to have lesser effects than approximating the variance.\\
Our work should be extended in two directions: to other domain
shapes that are of interest for the practitioners and to
3-dimensional data for high resolution medical imagery. A further
study of the asymptotics of the distribution of $\hat K(r)$ for
dependent point process models such as Markov or Cox processes
should also be achieved to inform on the power of our test.
\section{Proofs}
\subsection{ Proof of proposition \ref{prbi}} 
\begin{figure}
\centering \makebox{\includegraphics{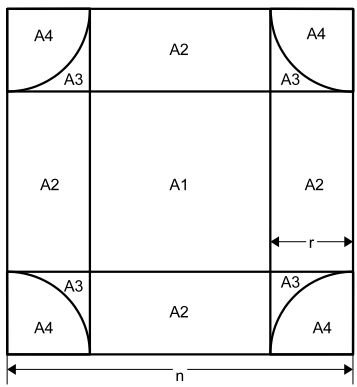}}
\caption{\label{f2}Zones in the square}
\end{figure}

Recall that $U$ and $V$ are two independent uniform variables on $A_n$.
The expectations of the Ripley statistics are
\begin{align*} \E\widehat K_{1,n}(r)&=\frac 1 {n^2\rho^2
}\E\left(\sum_{X_i\neq X_j \in
S} \1\{d(X_i,X_j)\leq r\} \right)\\
&=\frac {\E\left(N(N-1)\right)} {n^2\rho^2 } \E(\ \1\{d(U,V)\leq
r\})\\&=n^2 e_{r,n}.\\
\E\widehat K_{2,n}(r)&=n^2\E\left(\frac 1 {N(N-1) }\sum_{X_i\neq
X_j \in S} \1\{d(X_i,X_j)\leq r\} \right)\\&=n^2\P\left(N>1\right)
\E(\ \1\{d(U,V)\leq r\})\\&=n^2\left(1-e^{-\rho n^2}-\rho
n^2e^{-\rho n^2}\right) e_{r,n} .\end{align*} The following lemma
allows to conclude:
\begin{lem}\label{lemern}
$$e_{r,n}=\frac{\pi r^2}{n^2} -\frac {8r^3}{3n^3}+\frac {r^4}{2n^4}.$$
\end{lem}
{\it Proof:} We split $A_n$ into four parts to compute $e_{r,n}$:
\begin{eqnarray}
  e_{r,n} &=& \int_{\xi\in A_n^1}\int_{\eta\in A_n}\1\{d(\xi,\eta)\leq r\}\frac 1{n^4}d\xi
d\eta \label{in}\\
   &+&\int_{\xi\in A_n^2}\int_{\eta\in A_n}\1\{d(\xi,\eta)\leq r\}\frac 1{n^4}d\xi
d\eta  \label{ed}\\
   &+&\int_{\xi\in A_n^3}\int_{\eta\in A_n}\1\{d(\xi,\eta)\leq r\}\frac 1{n^4}d\xi
d\eta\label{te}\\
   &+&\int_{\xi\in A_n^4}\int_{\eta\in A_n}\1\{d(\xi,\eta)\leq r\}\frac 1{n^4}d\xi
d\eta\label{co}
\end{eqnarray}
where (see figure \ref{f2})\begin{itemize}
    \item  (interior) $A_n^1=$\{$\xi$, $\xi$ is at distance larger than $r$ from the boundary\}
    \item  (edge) $A_n^2=$\{$\xi$, $\xi$ is at distance less than $r$ from an edge, larger than $r$ from the others\}
    \item (two edges) $A_n^3=$\{$\xi$, $\xi$ is at distance less than $r$ from two edges and larger than $r$ from the corner\}
    \item (corner) $A_n^4=$\{$\xi$, $\xi$ is at distance less than $r$ from the corner\}
\end{itemize}

Note that $A_n^2$, $A_n^3$ and  $A_n^4$ are composed of four parts
that contribute identically. We establish formulas only for one of
these parts.
\begin{lem} \label{mdelt} Define function $g(x)=\arccos(x)-x\sqrt{1-x^2}$.\\If $\xi\in A_n^1$, $$ \int_{\eta\in
A_n}\1\,\{d(\xi,\eta)\leq r\}
d\eta=\pi r^2.$$\\
If $\xi\in A_n^2$, with $n-r<\xi_1<n$,  $x_1=\frac 1r (n-\xi_1)$,
$$ \int_{\eta\in A_n}\1\{d(\xi,\eta)\leq r\}
d\eta=r^2(\pi-g(x_1))$$ If $\xi\in A_n^3$, with $n-r<\xi_1<n$,
$n-r<\xi_2<n$ and $(x_1,x_2)=\frac 1r (n-\xi_1,n-\xi_2)$,
 $$ \int_{\eta\in A_n}\1\{d(\xi,\eta)\leq r\}
d\eta=r^2(\pi-g(x_1)-g(x_2)).$$ If $\xi\in A_n^4$, with
$n-r<\xi_1<n$, $n-r<\xi_2<n$ and $(x_1,x_2)=\frac 1r
(n-\xi_1,n-\xi_2)$,
 $$ \int_{\eta\in A_n}\1\{d(\xi,\eta)\leq r\}
d\eta=r^2\left(\frac{3\pi}{4}+x_1x_2-\frac{g(x_1)+g(x_2)}{2}\right).$$
\end{lem}\begin{figure}
\centering \makebox{\includegraphics{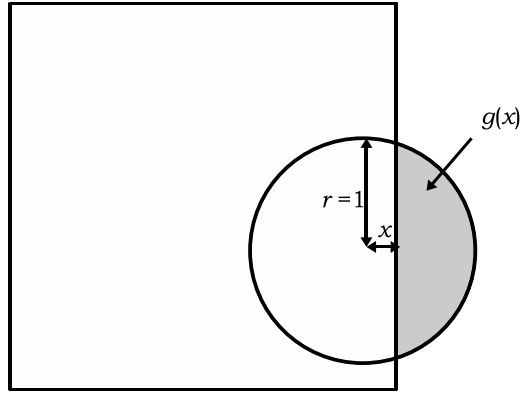}}
\caption{\label{f3}Geometrical interpretation of $g$}
\end{figure}
\noindent{\it Note:} Function $g(x)$ is the area of the part of
a ball of radius $1$ that lies outside the square when the ball
intersects one of its edges (see figure \ref{f3}).\\
{\it Proof.} For the
interior points $\xi \in A_n^1$, $B(\xi,r)\subset A_n$. \\
Let $\xi\in A_n^2$. We compute the area of $B(\xi,r)\cap A_n$.
\begin{eqnarray*}
  \int_{\eta\in A_n}\1\{d(\xi,\eta)\leq r\} d\eta&=&\frac {\pi r^2} 2
+2r^2\int_0^{x_1}\sqrt{1-t^2}dt \\
  &=&r^2\left(\pi -\arccos(x_1)+x_1\sqrt{1-x_1^2}\right)\\
  &=&r^2\left(\pi -g(x_1)\right).
\end{eqnarray*}
Note that $r^2g(x)$ is the part of the ball that lies out of the
square $A_n$ if the center is at distance $xr$ from the edge of the
square.
\\ Let $\xi\in A_n^3$. Here the ball intersects two edges of
the square and the area of $B(\xi,r)\cap A_n$ is
$$
\int_{\eta\in A_n}\1\{d(\xi,\eta)\leq r\} d\eta=r^2\left(\pi -
g(x_1)-g(x_2)\right).
$$
Let $\xi\in A_n^4$. Divide the ball into four quarters along axes
parallel to the coordinate axes. One of the quarter is inside the
square, two intersect the edges, leaving outside an area equal to
$(g(x_1)+g(x_2))/2$. The area of the intersection of the last
quarter with the square is $x_1x_2$ so that the area of
$B(\xi,r)\cap A_n$ is
$$
\int_{\eta\in A_n}\1\{d(\xi,\eta)\leq r\}
d\eta=r^2\left(\frac{3\pi} 4+x_1x_2 -\frac
{g(x_1)+g(x_2)}2\right).\quad \square
$$
{\it Proof of lemma \ref{lemern}(continued).} The left-hand side
of (\ref{in}) is $m(A_n^1)\pi r^2=\pi (n-2r)^2r^2$. Recall that
$A_n^2$ is composed of four parts that contribute identically. We
 integrate function $g$.
\begin{lem}$$
G(x)=\int_0^xg(u)du=x\arccos(x)-\sqrt{1-x^2}+\frac13(1-x^2)^{3/2}+\frac23.
$$
\end{lem}
{\it Proof.} Changing variables and integrating by
parts\begin{eqnarray*}
\int_0^x\arccos(u)du&=&-\int_{\pi/2}^{\arccos(x)}t\sin(t)dt\\&=&\left[t\cos
(t)\right]_{\pi/2}^{\arccos(x)}+\int_{\pi/2}^{\arccos(x)}\!\!\!\cos
(t)dt\\&=&x\arccos( x)-\sqrt{1-x^2}+1.\end{eqnarray*} Changing the
variable $v=\sqrt{1-u^2}$, we get
\begin{eqnarray*}
-\int_0^xu\sqrt{1-u^2}du&=&\int_1^{\sqrt{1-x^2}}v^2
dv=\frac13\left((1-x^2)^{3/2}-1\right).\quad\square\end{eqnarray*}
Then the contribution (\ref{ed}) is equal to
$$ 4r\int_r^{n-r}d\xi_2\int_0^1
r^2(\pi-g(x))dx=4 r^3(n-2r)(\pi-G(1))
=\left(4\pi-\frac83\right) r^3(n-2r). $$ We
consider $A_n^3$; the domain of integration  is symmetric in
$(x_1,x_2)$ so that  the contribution (\ref{te})  is equal
 to
 $$
    4r^4\!\!\int_0^1dx_1\int_{\sqrt{1-x_1^2}}^1(\pi
-2g(x_1))dx_2
   \! =\!4r^4\left(\pi\left(1-\frac{\pi}{4}\right)-2\int_0^1\!\!g(x_1)dx_1\!\!\int_{\sqrt{1-x_1^2}}^1
dx_2\right).
$$ From Lemma \ref{cintg},
\begin{eqnarray*}\int_0^1g(x_1)dx_1\int_{\sqrt{1-x_1^2}}^1dx_2
=G(1)-\int_0^1g(x_1)\sqrt{1-x_1^2}dx_1
=\frac23-\frac{\pi^2}{16}.\end{eqnarray*}
 so that
contribution (\ref{te})  is equal to $ \displaystyle
r^4\left(4\pi-\frac{\pi^2}{2}-\frac{16}3\right).$
\\We consider $A_n^4$; the contribution
(\ref{co}) is equal to\begin{eqnarray*}
4r^4\!\!\int_0^1\!\!\!\!dx_1\!\!\int_0^{\sqrt{1-x_1^2}}\!\!\left(\frac{3\pi}4
+x_1x_2 -g(x_1)\right)\!\!dx_2\!\!\!\!\!
&=&\!\!\!\!r^4\left(\frac{3\pi^2} {4}+\frac {1}2-4
\int_0^1\!\!\!\!g(x_1)\sqrt{1-x_1^2}dx_1\right)
   \\&=&\!\!\!\!r^4\left(\frac{\pi^2}2+\frac {1}2\right).\end{eqnarray*}
Gathering the four contributions, we get
\begin{eqnarray*} e_{r,n}&=&\frac {r^2}{n^2}\left(\pi
\left(1-\frac {2r}n\right)^2+\left(4\pi-\frac83\right) \frac
{r}n\left(1-\frac {2r}n\right)+\left(4\pi -\frac{29}6\right)\frac
{r^2}{n^2}\right)\\&=& \frac {r^2}{n^2}\left(\pi -\frac {8}3\frac
{r}n+\frac12\frac {r^2}{n^2}\right).\quad\square
\end{eqnarray*}
\subsection{  Proof of proposition \ref{pvar}}
We decompose the variance of $K_{s,A_n}(r) $ by conditioning the
variable with respect to the number $N$  of points in the sample.
Conditionally to $N$, $K_{s,A_n}(r) $  has the form of a
$U$-statistic. Then we apply the H\"{o}ffding decomposition to
this $U$-statistic. \\For $s=1,2$, we use the relation
$$\var(\widehat K_{s,A_n}(r))=\var\E (\widehat K_{s,A_n}(r)|N)+\E
\var(\widehat K_{s,A_n}(r)|N).$$  We first consider the conditional
expectation of $\widehat K_{s,A_n}(r)$.
\begin{eqnarray*} \E(\widehat K_{1,n}(r)|N)
&=&\frac 1 {n^2\rho^2 }\left(\sum_{i\neq j =1 }^N \E\,
\1\{d(X_i,X_j)\leq r\} \right) =\frac
{N(N-1)e_{r,n}}{n^2\rho^2},\\\E(\widehat K_{2,n}(r)|N) &=&\frac
{n^2}{N(N-1) }\sum_{i\neq j =1 }^N \E\, \1\{d(U_i,U_j)\leq r\}
=n^2e_{r,n}\1\{N>1\}.
\end{eqnarray*}
Because $N$ is a Poisson variable with intensity $\rho n^2$
\begin{eqnarray} \E N^2(N-1)^2&=&\E N(N-1)(N-2)(N-3)\nonumber\\&+&4\E
N(N-1)(N-2)+2\E N(N-1)\nonumber\\&=&\rho^4 n^8+4\rho^3 n^6+2\rho^2 n^4.\nonumber\\
\var N(N-1)&=&4\rho^3 n^6+2\rho^2 n^4.\label{N2}\end{eqnarray}
Then
\begin{eqnarray}\var\E(\widehat
K_{1,n}(r)|N) &=&\frac{(4\rho n^2+2)e^2_{r,n} }{\rho^2}\label{esck1}.\\
\var\E(\widehat K_{2,n}(r)|N)
&=&n^4\P\{N>1\}(1-\P\{N>1\})e^2_{r,n}\nonumber\\&=&n^4e^{-\rho
n^2}\left(1+\rho n^2\right)\left(1-e^{-\rho n^2}\left(1+\rho
n^2\right)\right)e^2_{r,n}.\label{esck2}
\end{eqnarray}We compute the conditional variances.  \begin{eqnarray*} \var(\widehat
K_{1,n}(r)|N) &=&\frac 1 {n^4\rho^4 }\var\left(\sum_{i\neq j =1
}^N h(X_i,X_j,r) \right) ,\\\var(\widehat K_{2,n}(r)|N) &=&\frac
{n^4}{N^2(N-1)^2 }\var\left(\sum_{i\neq j =1 }^N h(X_i,X_j,r)
\right).
\end{eqnarray*}Conditionally to $N$, the
locations of the points are independent and uniformly distributed
variables $U_i$ over $A_n$. We introduce the H\"{o}ffding
decomposition of the  $U$-statistic kernel $h$:
$$h(x,y,r)=h_1(x,r)+h_1(y,r)+h_2(x,y,r),
$$
where $h_1(x)=\E (h(U,V,r)|V=x)$, $(U,V)$ being two independent
uniform random variables on $A_n$.\\ Then $\E h_1(U,r)=0$ and $\E
(h_2(U,V,r)|U)=\E (h_2(U,V,r)|V)=0$, so that
\begin{eqnarray*}
\var h(U,V,r)&=&\var h_1(U,r)+\var h_1(V,r)+\var
h_2(U,V,r)\\&=&2\E h^2_1(U,r)+\var h_2(U,V,r). \end{eqnarray*}
From$$
\sum_{i\neq j = 1} ^N h(U_i,U_j,r)=2(N-1)\sum_{i = 1} ^N
h_1(U_i,r)+\sum_{i\neq j = 1} ^N h_2(U_i,U_j,r).
$$ we get
\begin{align*}
\var(\widehat K_{1,n}(r)|N)&=\frac {4(N-1)^2} {n^4\rho^4
}\var\left(\sum_{i= 1} ^Nh_1(U_i,r)\right)+\frac {1} {n^4\rho^4 }
\var\left(\sum_{i\neq j = 1}
^Nh_2(U_i,U_j,r)\right)\\
&=\frac {4N(N-1)^2} {n^4\rho^4 }\E h^2_1(U,r)+\frac {2}
{n^4\rho^4 } \sum_{i\neq j = 1} ^N\var h_2(U_i,U_j,r)\\
&=\frac {4N(N-1)^2} {n^4\rho^4 }\E h^2_1(U,r)+\frac {2N(N-1)}
{n^4\rho^4 } (\var h(U,V,r)-2\E h^2_1(U,r))\\
&=\frac {4N(N-1)(N-2)} {n^4\rho^4 }\E h^2_1(U,r)+\frac {2N(N-1)}
{n^4\rho^4 } \var h(U,V,r),
\end{align*} Now $\var
h(U,V,r)=e_{r,n}-e_{r,n}^2$ and using factorial moments of the
Poisson distribution
\begin{eqnarray}\E\ \var(\widehat K_{1,n}(r)|N)=\frac{4n^2}\rho\E
h_1^2(U,r)+\frac2{\rho^2}\left(e_{r,n}-e_{r,n}^2\right)\label{t1carre}.
\end{eqnarray}
Lemma \ref{h1carre} gives the exact value of $\E h_1^2(U,r)$. With
relations  (\ref{esck1}) and (\ref{t1carre}),  we get
\begin{eqnarray*}
\var(\widehat
K_{1,n}(r))&=&\frac{2e_{r,n}}{\rho^2}+\frac{4n^2e^2_{r,n}}{\rho
}+\frac{4n^2}{\rho }\E h_1^2(U_j,r)\\
  &=&\frac1{n^2}\left(\frac {2\pi
r^2}{\rho^2}+\frac {4\pi ^2r^4}{\rho}\right)\\&-&\frac
1{n^3}\left(\frac{16}3\frac {r^3}{\rho^2 }+\left( \frac {32\pi
}{3}+\frac{1024}{45}\right)\frac {r^ 5}\rho\right)\\&+& \frac
1{n^4}\left(\frac {r^4}{\rho^2}+\left( \frac {59\pi}{12}+\frac
{32}{9}\right)\frac {r^6 }{\rho}\right).
\end{eqnarray*}
Similarly
\begin{eqnarray*}
\var(\widehat K_{2,n}(r)|N)&=&\frac{\1\{N>1\}(N-2)}{N(N-1)}\E
h^2_1(U,r)+\frac {2n^4\ \1\{N>1\}}{N(N-1) } \var h(U,V,r),\\
\E\  \var(\widehat
K_{2,n}(r)|N)&=&\E\left(\frac{\1\{N>1\}(N-2)}{N(N-1)}\right)\!\!\E
h^2_1(U,r)\\&+&2n^4\E\left(\frac{\1\{N>1\}}{N(N-1)}
\right)\left(e_{r,n}-e_{r,n}^2\right).
\end{eqnarray*}
From this and relation  (\ref{esck2}),  we get\begin{eqnarray*}
\var(\widehat K_{2,n}(r)) &=&2n^4\E\left(\frac{\1\{N>1\}}{N(N-1)}
\right)\left(e_{r,n}-e_{r,n}^2\right)
\\&+&4n^4\E\left(\frac{\1\{N>1\}(N-2)}{N(N-1)}\right)\E
h_1^2(U_j,r)\\&+&n^4e^{-\rho n^2}\left(1+\rho
n^2\right)\left(1-e^{-\rho n^2}-\rho n^2e^{-\rho
n^2}\right)e^2_{r,n} .
\end{eqnarray*}We now apply the same decomposition to $\cov(\widehat
K_{1,n}(r),\widehat K_{1,n}(r')) $,\begin{equation}\label{tcvesp1}
    \cov(\E(\widehat K_{1,n}(r')|N),\E(\widehat K_{1,n}(r)|N))
=\frac{(4\rho n^2+2)e_{r',n}e_{r,n} }{\rho^2}.
\end{equation}
\begin{eqnarray*}
\cov(\widehat K_{1,n}(r'),\widehat K_{1,n}(r)|N)\!\! &=&\!\!\frac
{4(N-1)^2} {n^4\rho^4 }\cov\left(\sum_{i= 1}
^Nh_1(U_i,r'),\sum_{i= 1} ^Nh_1(U_i,r)\right)\\\!\!&+&\!\!\frac
{1} {n^4\rho^4 } \cov\left(\sum_{i\neq j = 1}
^Nh_2(U_i,U_j,r'),\sum_{i\neq j = 1}
^Nh_2(U_i,U_j,r)\right)\\\!\!&=&\!\!\frac {4N(N-1)(N-2)}
{n^4\rho^4 }\cov( h_1(U,r'),h_1(U,r))\\\!\!&+&\!\!\frac {2N(N-1)}
{n^4\rho^4 } \cov (h(U,V,r'),h(U,V,r))
.\end{eqnarray*}\begin{eqnarray*} \E\ \cov(\widehat
K_{1,n}(r'),\widehat K_{1,n}(r)|N)\!\!
\!\!&=&\!\!\!\!\frac{4n^2}\rho\cov(
h_1(U,r'),h_1(U,r))+\frac2{\rho^2}(e_{r,n}-e_{r',n}e_{r,n})  .
\end{eqnarray*}
To compute $\cov(h_1(U,r'),h_1(U,r))$, the square $A_n$ should now
be split into 16 different zones according to the 4 zones of the
preceding section with respect to $r$ and the 4 zones with respect
to $r'$. Because of inclusions, the actual number of zones to
consider is reduced to 9. The corresponding computation is easy in
the center zone, but can not be achieved in a close form in the
edge bands and in the corner. We consider the following zones:
\begin{itemize}
    \item  (interior) $A_n^{1,1}=$\{$\xi$, $\xi$ is at distance larger than $r'$ from the
    boundary\},
    \item  (interior-edge) $A_n^{1,2}=$\{$\xi$, $\xi$ is at distance between $r$ and $r'$ from an edge, larger than $r'$ from the
    others\},
    \item (edge) $A_n^{2,2}=$\{$\xi$, $\xi$ is at distance less than $r$ from an edge, larger than $r'$ from the
    others\},
    \item (corner) $A_n^{3,3}=$\{$\xi$, $\xi$ is at distance less than $r'$ from two
    edges\}.

\end{itemize}
Denoting  $x_1=\frac 1{r} (n-\xi_1)$ and $x'_1=\frac
1{r'} (n-\xi_1)$ we get
\begin{eqnarray*}
    h_1(X_j,r')h_1(X_j,r)\!\!\!\!&=&\!\!\!\!\!\!\left(\frac{\pi r'^2}{ n^2}-e_{r',n}\right)\!\!\left(\frac{\pi r^2}{ n^2}-e_{r,n}\right)\mbox{ on
}A_n^{1,1},\\
 \!\!&=&\!\!\!\!\!\!\left(\frac{ \pi r'^2}{ n^2}-e_{r',n}-\frac{ r'^2}{ n^2}g(x'_1)\right)\!\!\left(\frac{\pi r^2}{ n^2}-e_{r,n}\right)\mbox{ on
}A_n^{1,2},\\
\!\! \!\!&=&\!\!\!\!\!\!\left(\frac{ \pi r'^2}{
n^2}-e_{r',n}-\frac{ r'^2}{ n^2}g(x'_1)\right)\!\!\left(\frac{ \pi
r^2}{ n^2}-e_{r,n}-\frac{ r^2}{ n^2}g(x_1)\right)\mbox{ on
}A_n^{2,2}.
\end{eqnarray*}
 Denote $\displaystyle
b_{r,n}=\left(\pi -\frac{n^2}{ r^2}e_{r,n}\right)=
\frac{8r}{2n}-\frac{r^2}{2n^2}.$
$$\cov(h_1(X_j,r'),h_1(X_j,r))=C(A_n^{1,1})+C(A_n^{1,2})+C(A_n^{2,2})+C(A_n^{3,3})$$
\begin{eqnarray*}
C(A_n^{1,1})&=&\frac{r'^2r^2}{n^4}\left(1-\frac{2r'}{n}\right)^2b_{r',n}b_{r,n}\\
C(A_n^{1,2})&=&4\left(1-\frac{2r'}{n}\right)\frac{r'^3r^2}{n^5}b_{r,n}
\int_{r/r'}^1(b_{r',n}-g(x'_1))dx'_1\\
C(A_n^{2,2})&=&4\left(1-\frac{2r'}{n}\right)\frac{r^3r'^2}{n^5}\int_{0}^1(b_{r',n}-g(rx_1/r'))(b_{r,n}-g(x_1))dx_1.
\end{eqnarray*}The first integral may be expressed in terms of function $G$, the second integral is elliptic and has to be numerically evaluated; as the
integrand is bounded and very smooth this can be achieved without
difficulties. To compute the term $C(A_n^{3,3})$, we rewrite the
different values of function $h_1$ with the help of indicator
functions:
\begin{align*}
    h_{A1}(x,r)&=b_{r,n}\1\{ x_1\ge 1;x_2\ge 1 \}\\
    h_{A2}(x,r)&=(b_{r,n}-g(x_2))\1\{ x_1\ge 1;\ x_2< 1 \}+(b_{r,n}-g(x_1))\1\{ x_2\ge 1;\ x_1< 1 \}\\
    h_{A3}(x,r)&=(b_{r,n}-g(x_1)-g(x_2))\1\{ x_1< 1;\ x_2< 1 ;\ x_1^2+x_2^2\ge 1\}\\
    h_{A4}(x,r)&=(b_{r,n}-\pi/4+x_1x_2-(g(x_1)+g(x_2))/2)\1\{x_1^2+x_2^2<1 \}
\end{align*}
For $x'=\frac1{r'} (n-\xi_1,n-\xi_2)$
$$C(A_n^{3,3})=4\frac{r^2r'^4}{n^6}\int_0^1\int_0^1\sum_{i=1}^4h_{Ai}(r'x'/r,r)
\times\sum_{i=3}^4h_{Ai}(x',r')dx'_1dx'_2$$
and this integral also can be  numerically evaluated.\\
{\it Note}: the whole computation of this term of the covariance could be numerically achieved, but it is preferable to use an exact computation whenever it is possible.\\
The case of the covariance of $K_{2,n}(r)$ is analogous:
$$
    \cov(\E(\widehat K_{2,n}(r')|N),\E(\widehat K_{2,n}(r)|N))
    =n^4e^{-\rho n^2}\!\!\left(1+\!\!\rho n^2\right)\!\!(1-\!e^{-\rho
n^2}\!(1+\rho n^2))e_{r'\!,n}e_{r,n}.
$$
\begin{eqnarray*}
 \E\ \cov(\widehat K_{2,n}(r'),\widehat K_{2,n}(r)|N) \!\!\!\!&=&\!\!\!\!4n^4\E\left(\frac {
\1\{N>1\}(N-2)}{N(N-1)}\right)\cov(
h_1(U,r'),h_1(U,r))\\\!\!\!\!&+&\!\!\!\!2n^4\E\left(\frac{\1\{N>1\}}{N(N-1)}
\right)\left(e_{r,n}-e_{r',n}e_{r,n}\right) .\quad\square
\end{eqnarray*}

\subsection {Proof of Theorem \ref{tcl}.} We show that any linear
combination of the $K_{1,n}(r_t)$ is asymptotically normal.
 Let
$\Lambda=(\lambda_1, \ldots \lambda_d)$ be a vector of real
coefficients. Define $Z_1=\sum_{t=1}^d\lambda_tK_{1,n}(r_t)$. We
use the Bernstein blocks technique \citep{BE}: we divide the square $A_n$ into squares of side
$p$ with $p=o(n)$. These squares are separated by gaps of width
$2r_d$ so that the sums over couples of points in each square are
independent. The couples of points with at least  one point in the
gaps give a negligible contribution, so that the statistic $Z_1$
is equivalent to a sum of independent variables and asymptotically
normal.\\ Set $p=n^{1/4}$. Assume that the Euclidean division of
$n$ by $(p+2r_d)$ gives a quotient $a$ and a remainder $q$. For
$l=0,\ldots,a$, we define the segment $I_l=[(p+2r_d)l,
(p+2r_d)l+p-1]$. We order the set $\{0,\ldots,a\}^2$ by the
lexicographic order. To any integer $i$ such that $1\leq i \leq
k=(a+1)^2$, corresponds an element $(j_1,j_2)$ of this set; we
define the block $P_{i,n}=I_{j_l}\times I_{j_2}$ and
$Q=A_n\backslash\cup_iP_{i,n}$ the set of points that are  in none
of the $P_{i,n}$'s. For each block $P_{i,n}$ and $Q$, we define
the partial sums:
\begin{eqnarray*}
u_{i,n}&=&\frac 1{n\rho^{3/2}}\sum_{X_l\neq X_m \in
P_{i,n}}\sum_{t=1}^d
\lambda_t\ \1\{d(X_l,X_m)\leq r_t\},\\
v_{i,n}&=&\frac 1{n\rho^{3/2}}\sum_{X_l \in P_{i,n},X_m\in
Q}\sum_{t=1}^d
\lambda_t\ \1\{d(X_l,X_m)\leq r_t\}\\
w_n&=&\frac 1{n\rho^{3/2}}\sum_{X_l\neq X_m \in Q}\sum_{t=1}^d
\lambda_t\ \1\{d(X_l,X_m)\leq r_t\}.
\end{eqnarray*}
then
$$
n\sqrt\rho(Z_1-\E Z_1 )=\sum_{i=1}^{k}(u_{i,n}-\E
u_{i,n})+\sum_{i=1}^{k}(v_{i,n}-\E v_{i,n})+ w_n-\E w_n,
$$
 We show that the sum of the $u_{i,n}$ converges in distribution to a Gaussian variable
  and that the other term are negligible in ${\L}^2$.
We check the conditions of the following CLT adapted from
\cite{BD}.
\begin{theo}Let $(z_{i,n})_{0\leq i\leq k(n)}$ be an array of
 random variables satisfying
\begin{enumerate}
    \item There exists  $\delta> 0$ such that $\sum_{i=0 }^{k(n)}\E|z_{i,n}|^{ 2+\delta}$ tends to 0 as $n$ tends
    to infinity,
\item $\sum_{i=0 }^{k(n)}\var z_{i,n}$ tends to $\sigma^2$ as $n$
tends
    to infinity,
\end{enumerate}
then $\sum_{i=0}^{k(n)}z_{i,n}$ tends in distribution to $ {\cal
N}(0,\sigma^2)$ as $n$ tends to infinity.
\end{theo}
 To check Condition 1, we compute the fourth order moment of
$u_{i,n}-\E u_{i,n}$. Let $N_i$ be the number of points of $S$ that fall in
$P_{i,n}$. Define  $$f(x,y)=\sum_{t=1}^d \lambda_t\
(\1\{d(x,y)\leq r_t\}-e_{r,p})=\sum_{t=1}^d \lambda_t\
h(x,y,r_t)$$
$$
\E((u_{i,n}-\E u_{i,n})^4|N_i)=\frac 1 {n^4\rho^6
}\E\left(\sum_{l\neq m = 1} ^{N_i} f(U_l,U_m)\right) ^4$$ Denote
$f_1$ and $f_2$ the decomposing functions  of $f$:\\
$\E(f_1(U_l))=0$,
$\E(f_1(U_l)f_2(U_l,U_m))=\E(f_1(U_m)f_2(U_l,U_m))=0$, for $U_l$
and $U_m$ two independent uniform variables on $P_{i,n}$.
$$
  \sum_{l\neq m = 1}
^{N_i} f(U_l,U_m)= 2(N_i-1)\sum_{l= 1} ^{N_i} f_1(U_l)+\sum_{l\neq
m = 1} ^{N_i} f_2(U_l,U_m). $$   Note that $|h_1(x,r)|\leq\pi
r^2p^{-2}$ so that $f_1$ is bounded by $Cp^{-2}$.\\ Define
$M_1=\E\left(\sum_{l = 1} ^{N_i} f_1(U_l)\right) ^4 . $ Then $
M_1=N_iE(f_1^4(U))+6N_i(N_i-1)E(f_1^2(U))^2$ and
$$\E (N_i-1)^4M_1= { O}(1).$$
Define $M_2= \E\left(\sum_{l\neq m = 1} ^{N_i} f_2(U_l,U_m)\right)
^4 $.
 Because $f_2$ is zero
mean with respect to one coordinate, only the products where
variables appear at least two times contribute.
  \begin{eqnarray*}M_2&=
&\!\!\!8\sum_{l\neq m = 1} ^{N_i} \E f_2^4(U_l,U_m)
+16\!\!\!\sum_{l\neq m\neq u = 1} ^{N_i}\!\!\!\E
f_2^2(U_l,U_u)f_2^2(U_m,U_u)\\&+&\!\!\!32\sum_{l\neq m\neq u = 1} ^{N_i}\E f_2^2(U_l,U_m)f_2(U_m,U_u)f_2(U_l,U_u)\\
&+&4\sum_{l\neq m \neq u\neq v= 1} ^{N_i}\E
 f_2^2(U_l,U_m)f_2^2(U_u,U_v)\\
&+&16\sum_{l\neq m \neq u\neq v= 1} ^{N_i}\E
 f_2(U_l,U_m)f_2(U_m,U_u)f_2(U_u,U_v)f_2(U_v,U_l).
 \end{eqnarray*} Because $f_2$ is bounded,
$\displaystyle
\E M_2=O(\E N_i(N_i-1)(N_i-2)(N_i-3)) =O(p^{8})
$, so that
$$
\sum_{i=0 }^{k}\E(u_{i,n}-\E u_{i,n})^4=O(p^{6}n^{-2}).
$$
As $p=n^{1/4}$, we get condition 1. \\
To check condition 2, note that the vector
$(K_{1,P_i}(r_1),\ldots,K_{1,P_i}(r_d))$ has a covariance matrix
$\Sigma_p$ defined by Proposition \ref{pvar} by substituting $p$
to $n$ in the expressions. The
$u_{i,n}=\frac{p^2\sqrt\rho}n\sum_{t=1}^d\lambda_t(K_{1,P_i}(r_t)-\E
K_{1,P_i}(r_t))$ are i.i.d variables with variance equal to $
\frac{p^4\rho}{n^2}\Lambda^t\Sigma_p\Lambda$. But
$p^2\rho\Sigma_p$ tends to $\Sigma$ as $p$ tends to infinity and
$$\sum_{i=0 }^{k}\var u_{i,n}=\frac{kp^4\rho}{n^2}\Lambda^t\Sigma_p\Lambda\longrightarrow\Lambda^t\Sigma\Lambda$$
so that $ \sum_{i=1}^ku_{i,n}$ tends in distribution to ${\cal
N}(0,\Lambda^t\Sigma\Lambda)$.\\
Note that the $v_{i,n}$ are $k$ independent variables. Denote
$N_{i,r_d}$ the number of points $X_l$ in the  boundary region
$P_{i,r_d}$ of $P_{i,n}$ such that the ball $ B(X_l,r_d)$
intersects $Q$ and let $D(X_l)$ denote this intersection. Note
that $$\E N_{i,r_d}=\rho m(P_{i,r_d})\leq Cpr_d.$$
\begin{eqnarray*}
  \var v_{i,n}&\leq& \frac C {n^2 }\E\left(\sum_{l= 1} ^{N_{i,r_d}}\sum_{m = 1}
^{N_{Q}} \1\{X_m\in D(X_l)\}\right) ^2
  \leq \frac {C} {n^2 }(T_1+T_2),\end{eqnarray*}
 where
 \begin{eqnarray*}
  T_1&=&\E\sum_{l= 1} ^{N_{i,r_d}}\sum_{m = 1} ^{N_{Q}} \sum_{u = 1}
^{N_{Q}}\1\{X_m\in D(X_l)\}\1\{X_u\in D(X_l)\}
\\T_2&=&  \E\sum_{l= 1}^{N_{i,r_d}}\sum_{m = 1} ^{N_{i,r_d}} \sum_{u =
1}^{N_{Q}}\1\{X_u\in D(X_l)\cap D(X_m)\}.
\end{eqnarray*}
\begin{eqnarray*}
T_1 &\leq&\E N_{i,r_d}\E N_{Q}^2\P^2\{X_m\in D(X_l)|X_m\in Q\}\\&\leq&  \rho ^3 m(P_{i,r_d})(m^2(Q)+m(Q))\left(\frac{\pi r_d^2}{2m(Q)}\right)^2=O(p).\\
T_2&=&\E\sum_{l= 1}^{N_{i,r_d}}\sum_{m = 1} ^{N_{i,r_d}} \sum_{u =
1}^{N_{Q}}\1\{X_m\in B(X_l,2r_d)\}\1\{X_u\in D(X_l)\cap D(X_m)\}\\&
\leq&\E N_{i,r_d}^2\P\{X_m\in B(X_l,r_d)|X_m\in P_{i,r_d}\}\E
N_{Q}\P\{X_u\in D(X_l)|X_u\in Q\}\\&\leq&  \rho
^3(m^2(P_{i,r_d})+m(P_{i,r_d}))\left(\frac{\pi
r_d^2}{m(P_{i,r_d})}\right)m(Q) \left(\frac{\pi
r_d^2}{2m(Q)}\right)=O(p).
\end{eqnarray*}
and
$\var\left(\sum_{i=1}^kv_{i,n}\right)=O\left(kp/n^2\right)=O\left(
p^{-1}\right)$, so that this sum is negligible in ${\L}^2$. Similarly
\begin{eqnarray*}
\var(w_{n}) &\leq& \frac {C} {n^2 }\E\left(\sum_{l\neq m= 1}
^{N_{Q}}
 \1\{X_m\in B(X_l,r_d)\}\right) ^2
 \le\frac {C} {n^2 }(T_1+T_2).
\end{eqnarray*} where
\begin{eqnarray*}
 T_1&=&\E\sum_{l= 1} ^{N_{Q}}\sum_{m = 1}^{N_{Q}} \1\{X_m\in
 B(X_l,r_d)\}\\
&\le&\E N_{Q} (N_{Q}-1)\P\{X_m\in B(X_l,r_d)|X_m\in Q\} \le
m^2(Q)\frac{\pi r_d^2}{m(Q)}.\\
 T_2&=&\E\sum_{l= 1} ^{N_{Q}}\sum_{m = 1} ^{N_{Q}} \sum_{u = 1} ^{N_{Q}}\1\{X_m\in B(X_l,r_d)\}\1\{X_u\in
 B(X_l,r_d)\}\\
&\le& \E N_{Q}^2(N_Q-1)\P^2\{X_m\in B(X_l,r_d)|X_m\in Q\}\\&
\le&(m^3(Q)+2m^2(Q)) \left(\frac{\pi r_d^2}{m(Q)}\right)^2.
\end{eqnarray*}
Then $\var(w_{n}) =O\left( m(Q)/n^2\right)=O\left( p^{-1}\right) $ and $w_n$ is negligible in ${\L}^2$.\\
 Consider now $K_{2,n}(r)$. Define
$Z_2=\sum_{t=1}^d\lambda_tK_{2,n}(r_t)=A_{N,n}Z_1$ where
$A_{N,n}=\frac{n^4\rho^2}{N(N-1)}$.  We have $\E(A_{N,n}^{-1})=1$
and from (\ref{N2}), $\displaystyle \var(A_{N,n}^{-1})=\frac
4{n^2\rho}+\frac 2{n^4\rho^2}$. \\For $\delta>0$, the Markov
inequality gives
$$\P(|A_{N,n}^{-1}-1|>\delta)\le
\frac{\var(A_{N,n}^{-1})}{\delta^2}.$$ Then, with
$\delta=n^{-1/4}$
$$\sum_{n=1}^\infty\P(|A_{N,n}^{-1}-1|>n^{-1/4})<\sum_{n=1}^\infty\frac 4{n^{3/2}\rho}+\frac 2{n^{7/2}\rho^2}<\infty.$$
From the Borel-Cantelli lemma, we get that $A_{N,n}^{-1}$
converges a.s. to 1.  By the Slutsky lemma, $A_{N,n}Z_1$ converges in distribution to ${\cal
N}(0,\Lambda^t\Sigma\Lambda).\quad\square$
\subsection{Computation of $\E h_1^2(U,r)$ }
\begin{lem}\label{h1carre}
\begin{eqnarray*}
\E h_1^2(U,r)
        &=&\frac{r^5}{n^5}\left(\frac 8 3\,\pi -{\frac {256}{45}}\right)
        +\frac{r^6}{n^6}\left(\frac {11}{48}\,\pi -\frac {56}
{9}\right)+\frac 8 3\frac{r^7}{n^7}-\frac 1 4\frac{r^8}{n^8}.
 \end{eqnarray*}
\end{lem}
\noindent{\it Proof: }From the computation of the bias, denoting
$x_i=\frac 1r (n-\xi_i)$, we get
\begin{eqnarray*}
    h_1(\xi,r)&=&\frac{\pi r^2}{ n^2}-e_{r,n}\mbox{ on
}A_n^1\\
 &=&\frac{r^2}{n^2}(\pi-g(x_1))-e_{r,n}\mbox{ on
}A_n^2\\
 &=&\frac{r^2}{n^2}(\pi-g(x_1)-g(x_2))-e_{r,n}\mbox{ on
}A_n^3\\
&=&\frac{r^2}{n^2}\left(\frac{3\pi }{4}+x_1x_2-\frac
{g(x_1)+g(x_2)}2\right)-e_{r,n}\mbox{ on }A_n^4
\end{eqnarray*}
\begin{eqnarray}
\E(h_1(X_j,r))^2\!\!&=&\!\!\pi^2\left(1-\frac{2r}{n}\right)^2
\frac{r^4}{n^4}-e^2_{r,n}+4\left(1-\frac{2r}{n}\right)
\frac{r^5}{n^5}T_1+4\frac{r^6}{n^6}(T_2+T_3)\nonumber
 \\
T_1 &=&\!\!\int_0^1(\pi-g(x_1))^2dx_1\label{l2}\\
T_2 &=&\!\!\int_0^1dx_1\int_{\sqrt{1-x_1^2}}^1(\pi-g(x_1)-g(x_2))^2dx_2\label{l3}\\
T_3
&=&\!\!\int_0^1\!\!dx_1\!\!\int_0^{\sqrt{1-x_1^2}}\!\!\!\left(\frac{3\pi
}{4}+x_1x_2-\frac {g(x_1)+g(x_2)}2\right)^2\!\!dx_2\label{l4}.
\end{eqnarray}
To compute these three terms, we need integral computations on
function $g$.
\begin{lem}\label{intarc}For $n\geq 1$,
\begin{eqnarray*}
I_n&=&\int_0^{1}u^{2n-1}\arccos (u)
du=\frac{\pi(2n)!}{n2^{2n+2}(n!)^2}.
\\
J_n&=&\int_0^{1}u^{2n}\sqrt{1-u^2}du=-(2n+2)I_{n+1}+2nI_{n}.
\end{eqnarray*}
\begin{eqnarray}
\label{acos}
    \int_0^1\sqrt{1-u^2}\arccos( u)du&=&\frac{\pi^2} {16}+\frac14.
\\\label{acos2} \int_0^1\sqrt{1-u^2}\arccos^2(u)du&=&\frac{\pi^3}
{48}+\frac{\pi} {4}.
\end{eqnarray}
\end{lem}
\noindent{\it Note:} in the following, we use $I_1=\pi/8$,
$I_2=3\pi/64$, $J_1=\pi/16$ and $J_2=\pi/32$.
\begin{lem}\label{cintg}\begin{eqnarray}\int_0^1g(u)\sqrt{1-u^2}du&=&\frac{\pi^2}
{16}.\label{racg}\\
    \int_0^1g^2(u)du &=&\frac{2\pi}{3}-\frac{64}{45}.\label{g2}\\
\int_0^1g^2(u)\sqrt{1-u^2}du &=&\frac{\pi^3}{48}.
\label{racg2}\\\int_0^1g(u) \ G\left(\sqrt{1-u^2}\right)du&=&\frac{\pi^3}{96}-\frac{5\pi}{48}
+\frac{4}{9}.\label{gG}
    \end{eqnarray}
 \end{lem}
\noindent Proofs are postponed in the appendix. Using these lemmas,
we get
  \begin{eqnarray}\label{tA1}
    T_1&=&\pi^2-2\pi G(1)+\int_0^1g^2(x_1)dx_1=\pi^2-\frac{64}{45} -\frac{2\pi} 3.\\
   T_2 &=& \pi^2\left(1-\frac{\pi}{4}\right) -4\pi\int_0^1g(x_1)dx_1\int_{\sqrt{1-x_1^2}}^1dx_2
    +2\int_0^1g^2(x_1)dx_1\int_{\sqrt{1-x_1^2}}^1dx_2\nonumber\\&&+2\int_0^1g(x_1)dx_1\int_{\sqrt{1-x_1^2}}^1g(x_2)dx_2.\nonumber
\end{eqnarray} From the computation of the bias,
$\displaystyle
  -4\pi\int_0^1g(x_1)dx_1\int_{\sqrt{1-x_1^2}}^1dx_2 =
  -\frac{8\pi}3+\frac{\pi^3}{4}.
$ \\From (\ref{g2}),  (\ref{racg2}) and (\ref{gG}), we get
$$
2\int_0^1g^2(x_1)dx_1\int_{\sqrt{1-x_1^2}}^1dx_2 =2\int_0^1g^2(x_1)dx_1-2\int_0^1\sqrt{1-x_1^2}g^2(x_1)dx_1
=\frac{4\pi} {3}-\frac{128}{45}-\frac{\pi^3} {24}.
$$
$$  2\int_0^1g(x_1)dx_1\int_{\sqrt{1-x_1^2}}^1
  g(x_2)dx_2= 2G^2(1)-2\int_0^1g(x_1)G\left(\sqrt{1-x_1^2}\right)dx_1 =-\frac{\pi^3}{48}+\frac{5\pi}{24}.
$$
Adding these results, we obtain
\begin{equation}\label{tA3}
T_2
    =-\frac{\pi^3}{16}+\pi^2-\frac{9\pi}{8}-\frac{128}{45}.\end{equation}
To compute $T_3$, we write
\begin{eqnarray*}
T_3 &=&\!\!\frac{9\pi ^3}{64}
+\int_0^1\!\!\!\!x_1^2dx_1\int_0^{\sqrt{1-x_1^2}}\!\!\!\!\!x_2^2dx_2
-\frac{3\pi
}{2}\int_0^1\!\!g(x_1)dx_1\int_0^{\sqrt{1-x_1^2}}\!\!\!\!\!\!\!\!dx_2
\\&+&\frac12\int_0^1g^2(x_1)dx_1\int_0^{\sqrt{1-x_1^2}}\!\!\!\!\!\!\!\!dx_2
    +\!\!\frac{3\pi }{2}\int_0^1\!\!\!\!x_1dx_1\int_0^{\sqrt{1-x_1^2}}\!\!\!\!\!\!\!\!x_2dx_2
    \\&+&\frac12\int_0^1\!\!\!\!\!g(x_1)dx_1\int_0^{\sqrt{1-x_1^2}}\!\!\!\!\!\!\!\!g(x_2)dx_2
    -2\int_0^1x_1g(x_1)dx_1\int_0^{\sqrt{1-x_1^2}}\!\!\!\!\!\!\!\!\!\!x_2dx_2.
\end{eqnarray*}
$ \displaystyle\int_0^1x_1^2dx_1\int_0^{\sqrt{1-x_1^2}}x_2^2dx_2 =
  \frac13\int_0^1x_1^2(1-x_1^2)\sqrt{1-x_1^2}dx_1=\frac13(J_1-J_2)=\frac{\pi}{96}.
$ \\
From (\ref{racg}), $ \displaystyle-\frac{3\pi
}{2}\int_0^1g(x_1)dx_1\int_0^{\sqrt{1-x_1^2}}dx_2=-\frac{3\pi
^3}{32}$.\\
From (\ref{racg2}), $ \displaystyle
\frac12\int_0^1g^2(x_1)dx_1\int_0^{\sqrt{1-x_1^2}}dx_2=-\frac{\pi^3}{96}.
$ \\ $\displaystyle \frac{3\pi
}{2}\int_0^1x_1dx_1\int_0^{\sqrt{1-x_1^2}}x_2dx_2  =
  \frac{3\pi }{4}\int_0^1x_1(1-x_1^2)dx_1=\frac{3\pi }{16}.
$
\\From (\ref{gG}), $\displaystyle
\frac12\int_0^1g(x_1)dx_1\int_0^{\sqrt{1-x_1^2}}g(x_2)dx_2=\frac{\pi^3}{192}-\frac{5\pi}{96}
+\frac{2}{9}. $\\
$\displaystyle
    -2\int_0^1x_1g(x_1)dx_1\int_0^{\sqrt{1-x_1^2}}x_2dx_2
    = \int_0^1\left(x_1^3-x_1\right)g(x_1)dx_1=-\frac{3\pi} {64}.
$\\
Adding these results, we get
\begin{equation}\label{tA4}
T_3 =\frac{\pi
        ^3}{16}+\frac{19\pi}{192}+\frac{2}{9}.
 \end{equation}
Gathering (\ref{tA1}), (\ref{tA3}) and (\ref{tA4}) gives the result.
 $\square$

\appendix

\section{Integration lemmas}
\subsection{Proof of Lemma \ref{intarc}} Integrating by parts
$$\int_0^{1}u^{2n-1}\arccos (u)
du=\int_0^{\pi/2}t\cos^{2n-1}(t)\sin (t)dt=\frac
1{2n}\int_0^{\pi/2}\cos ^{2n}(t)dt.$$ Using De Moivre formula
$$ \cos
^{2n}(t)=\frac{1}{2^{2n}}\left(2\cos(2nt)+2\left(\begin{array}{c}2n
\\ 1 \end{array}
\right) \cos(2(n-1)t)+\cdots+\left(\begin{array}{c} 2n\\n
\end{array} \right)\right).$$
Only the last  term gives a non zero integral, giving the result
for $I_n$.
\begin{eqnarray*}
J_n&=&\int_0^{1}(u^{2n+2}-u^{2n})(-(1-u^2)^{-1/2})du\\&=&\left[(u^{2n+2}-u^{2n})\arccos
(u)\right]_0^1-\int_0^{1}((n+2)u^{2n+1}-nu^{2n-1})\arccos (u) du
\end{eqnarray*}and the term under brackets is zero, giving the result.
\begin{eqnarray*}
\int_0^1\sqrt{1-u^2}\arccos(u)du&=&\int_0^{\pi/2}t\sin^2
(t)dt=\int_0^{\pi/2}\frac{t}2-\frac{t\cos (2t)}2dt
\\&=&\frac{\pi^2}{16}-\left[\frac{t\sin(2t)}4\right]_0^{\pi/2}+\int_0^{\pi/2}\frac{\sin(2t)}4dt=\frac{\pi^2} {16}+\frac14.\\
\int_0^1\sqrt{1-u^2}\arccos^2(u)du&=&\int_0^{\pi/2}t^2\sin^2(t)dt=\int_0^{\pi/2}\frac{t^2}2-\frac{t^2\cos(2t)}2dt
\\&=&\frac{\pi^3} {48}-\left[\frac{t^2\sin(2t)}4\right]_0^{\pi/2}+\int_0^{\pi/2}\frac{t\sin(2t)}2dt
\\&=&\frac{\pi^3} {48}-\left[\frac{t\cos
(2t)}4\right]_0^{\pi/2}+\int_0^{\pi/2}\frac{\cos
(2t)}4dt=\frac{\pi^3} {48}+\frac{\pi} {8}.\quad\square
\end{eqnarray*}
\newpage
\subsection{Proof of lemma \ref{cintg}}
Equation (\ref{racg}) follows from equation (\ref{acos}).\\
 Write $g^2(u)=\arccos^2
(u)+u^2-u^4-2u\sqrt{1-u^2}\arccos(u)$ and
\begin{eqnarray*}\int_0^1\arccos^2(u)du
&=&\int_0^{\pi/2}t^{2}\sin( t)dt=-\left[ t^{2}\cos(
t)\right]_0^{\pi/2}+2\int_0^{\pi/2}t\cos (t)dt\\&=&2\left[ t\sin(
t)\right]_0^{\pi/2}+2\int_0^{\pi/2}\sin(t)dt=\pi-2,\\
\int_0^1(u^2-u^4)du &=&\frac13-\frac15=\frac2{15}.\end{eqnarray*}
\begin{eqnarray*}\int_0^1u\sqrt{1-u^2}\arccos(u)du
&=&\int_0^{\pi/2}t\cos(t)\sin^{2}( t)dt\\&=&\left[\frac t3\sin^3(
t)\right]_0^{\pi/2}-\frac13\int_0^{\pi/2}\sin^3 (t)dt\\&=&\frac\pi
6-\frac13\int_0^{\pi/2}\sin (t)dt+\frac13\int_0^{\pi/2}\cos^2
(t)\sin (t)dt\\&=&\frac\pi 6-\frac13-\frac19\left[ \cos^3(
t)\right]_0^{\pi/2}=\frac\pi 6-\frac29.\end{eqnarray*}
Collecting the three parts yields to (\ref{g2}).
\begin{eqnarray*}
\int_0^1g^2(u)\sqrt{1-u^2}du &=&\int_0^1\sqrt{1-u^2}\arccos^2(u)du\\
&&-2\int_0^1(u-u^3)\arccos(u)du+\int_0^1\sqrt{1-u^2}(u^2-u^4)du\\
&=&\frac{\pi^3}{48}+\frac{\pi} {8}-2\left(\frac\pi{8}-\frac{3\pi}
{64}\right)+\frac{\pi} {16}-\frac{\pi}
{32}=\frac{\pi^3}{48}.
\end{eqnarray*}
  Write $\displaystyle G\left(\sqrt{1-x^2}\right)=\sqrt{1-x^2}\left(\frac\pi 2-\arccos(x)\right)+\frac{x^{3}}3-x+\frac23
$
\begin{eqnarray*}
   \int_0^1g(x)G\left(\sqrt{1-x^2}\right)dx&=&\int_0^1\sqrt
{1-x^{2}}\left(\frac{\pi}2-\arccos(x)\right)\arccos(x)dx\\&&-\int_0^1(x-x^3)\left(\frac{\pi}2-\arccos(x)\right)dx\\
&&+\int_0^1\left(\frac{x^3}{3}-x+\frac{2}{3}\right)\arccos(x)dx\\
&&+\int_0^1\left(-\frac{x^4}{3}+x^2-\frac{2x}{3}\right)\sqrt{1-x^2}dx
\\&=&\frac{\pi^3}{96}-\frac{5\pi}{48} +\frac{4}{9}.\quad \square\end{eqnarray*}

\end{document}